\documentclass[preprint,3p,twocolumn]{elsarticle}

\usepackage{xfrac}
\usepackage{caption}
\usepackage{subfig}
\usepackage{graphicx}

\usepackage{verbatim}
\usepackage[percent]{overpic}
\pdfoutput=1 
\usepackage{color}
\usepackage{siunitx}
\usepackage[T1]{fontenc}
\DeclareSIUnit[number-unit-product = {}]{\inchQ}{\textquotedbl}

\DeclareSIUnit[number-unit-product = {\thinspace}]{\inch}{in}

\usepackage{amssymb}

\journal{NIMA}

\newcommand{\iso}[2]{$^{#2}$#1}

\newcommand{\state}[3]{
    \ifnum #2=1
        $#1^{#3}$
    \else
        #1/#2$^{#3}$
    \fi
}

\begin{document}

\begin{frontmatter}

\title{Neutron detection efficiency of the Neutron dEtector with Xn Tracking (NEXT) }


\author[mymainaddress]{S.~Neupane\corref{mycorrespondingauthor}}
\cortext[mycorrespondingauthor]{Corresponding author}
\ead{sneupan4@vols.utk.edu}
\author[mymainaddress]{J.~Heideman}
\author[mymainaddress,ORNLaddress]{R.~Grzywacz}

\author[mymainaddress]{J.~Hooker}
\author[mymainaddress]{K.L.~Jones}
\author[mymainaddress]{N. Kitamura}
\author[mymainaddress]{C.R.~Thornsberry}

\author[UTKNEaddress]{L.H.~Heilbronn}

\author[TTUaddress]{M.M.~Rajabali}
\author[OUaddress]{Y. Alberty-Jones}
\author[OUaddress]{J. Derkin}
\author[OUaddress]{T. Massey}
\author[OUaddress]{D. Soltesz}

\address[mymainaddress]{Department of Physics and Astronomy, University of Tennessee, Knoxville, Tennessee 37996 USA}
\address[UTKNEaddress]{Department of Nuclear Engineering, University of Tennessee, Knoxville, Tennessee 37996 USA}
\address[ORNLaddress]{Physics Division, Oak Ridge National Laboratory, Oak Ridge, Tennessee 37831 USA}
\address[TTUaddress]{Department of Physics Tennessee Technological University, Cookeville, Tennessee, 38505, USA}
\address[OUaddress]{Department of Physics \& Astronomy, Ohio University, Athens, Ohio, 45701 USA}
\begin{abstract}

An efficient neutron detection system with good energy resolution is required to correctly characterize decays of neutron-rich nuclei where $\beta-$delayed neutron emission is a dominant decay mode. The Neutron dEtector with Xn Tracking (NEXT) has been designed to measure $\beta$-delayed neutron emitters. By segmenting the detector along the neutron flight path, NEXT reduces the associated uncertainties in neutron time-of-flight measurements, improving energy resolution while maintaining detection efficiency. Detector prototypes are comprised of optically separated segments of a neutron-gamma discriminating plastic scintillator coupled to position-sensitive photomultiplier tubes.  The first performance studies of this detector showed that high intrinsic neutron detection efficiency could be achieved while retaining good energy resolution. The results from the efficiency measurements using neutrons from direct reactions are presented.

\end{abstract}

\begin{keyword}
Neutron emission, neutron detection, efficiency

\end{keyword}

\end{frontmatter}



\section{Introduction}

Neutron spectroscopy is an important tool to study the nuclear properties of exotic, neutron-rich nuclei \cite{dimitriou2021development, nakamura2017exotic, pfutzner2012radioactive}. 
Beta-delayed neutron emission \cite{roberts1939further} is a significant
or even dominant decay channel for these nuclei. Radioactive ion
beam facilities continue to increase access to very neutron-rich nuclei  \cite{doi:10.1080/10619127.2013.855558}. Improvements in the quality of neutron detection
techniques must continue to fully exploit their discovery potential. Neutron-energy measurements are required to probe nuclear
structure effects, and these detection systems need to be capable of measuring neutrons with an
energy range of 100~keV to 10~MeV as efficiently as possible.

The Neutron dEtector with Xn Tracking (NEXT), an optically segmented plastic scintillator neutron time-of-flight (TOF) detector array, has been designed to study the $\beta$-delayed neutron emission. The detector segmentation enables high-precision measurement of neutron TOF and the position interaction, which improves the energy resolution without sacrificing neutron detection efficiency. With the interaction position localization, individual detector units can be placed closer to the source or target, making the detector array efficient and cost-effective without compromising the necessary energy resolution. NEXT can also be used in direct reactions studies using $(d,n)$, $(p,n)$, and $(\alpha,n)$. The neutron-gamma (n-$\gamma$) discrimination capability of NEXT is critical in reactions experiments where the gamma-ray background would otherwise be limiting. The NEXT concept and results from proof-of-principle tests are described by Heideman et al. \cite{heideman2019conceptual}. This publication will detail efficiency measurements with neutrons produced by \iso{Al}{27}(d,n) reactions and comparisons to GEANT4 simulations.

\section{Detector Prototypes}
\begin{figure}[!t]
\centering
\includegraphics[width=0.47\textwidth]{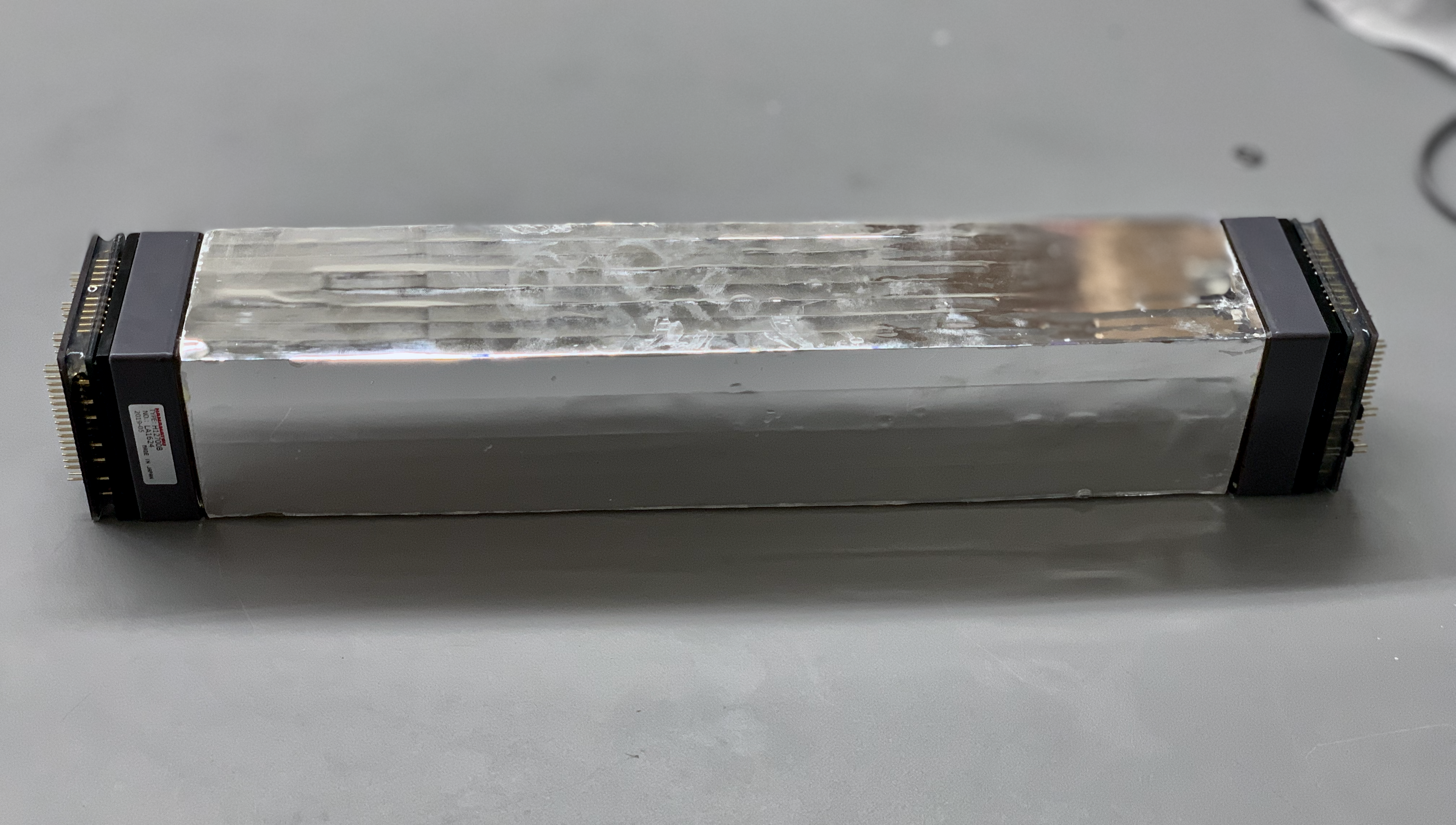}
	\caption{A typical NEXT prototype. The segmented plastic scintillator is coupled to Hamamatsu H12700 position-sensitive PMTs on either side for the light readout.}
\label{fig:prototype}
\end{figure}

A typical NEXT prototype has 8$\times$4 segments, where the higher segmentation is along the neutron flight path. Each segment is a bar of plastic scintillator developed by Eljen Technologies \cite{eljen}. The segments are 6~mm thick, 12.5~mm wide, and 254~mm long, making the active scattering volume of the entire detector module 48$\times$50.8$\times$254~mm$^3$. A fully assembled detector can be seen in Figure \ref{fig:prototype}. Each scintillator bar is optically separated from the others using 3M™ Enhanced Specular Reflector(ESR) \cite{3MESR}. The details of the conceptual design of the NEXT prototype can be found in  \cite{heideman2019conceptual}.

 Two different types of plastic scintillators manufactured by Eljen Technologies, EJ-200 and EJ-276, were used to construct the prototypes. The EJ-276 is a n-$\gamma$ discriminating plastic, while EJ-200 does not have any n-$\gamma$ discrimination capabilities, but it has about 15\% higher light yield than EJ-276. The prototypes are designated by the plastic type used and their length; for instance, EJ276-10 is made of EJ-276 plastic scintillator and 10~in long.
 
For the light readout, a  multi-anode photomultiplier tube (MAPMT), Hamamatsu H12700 \cite{HMPMT} was used on either side of the detector module. Each MAPMT has 8$\times$8 pixels, each of them has dimension 6$\times$6~mm$^2$, which makes an active window of 48$\times$48~mm$^2$. The signals from the MAPMT are read out using Anger Logic boards, designed and manufactured by Vertilon Corporation \cite{VERTILON}. The Anger Logic board, an array of matched resistors, reduces the number of readout channels to five channels per MAPMT. The position information is calculated using four signals from the corners of the resistor network using the Anger Logic algorithm \cite{anger1964scintillation}. The neutron interaction position inside the detector is used to correct the neutron TOF. The fifth signal is the dynode signal, common to one side of the detector, which carries the timing and pulse-shape discrimination information.  The five signals from each Anger Logic board are recorded using 16-bit, 250 MHz Pixie-16 digitizers developed by XIA  LLC \cite{pixie16}.

\section{Prototype Efficiency}
\subsection{$^{27}$Al(d,n)Measurement}

The neutron detection efficiency measurement of different NEXT prototypes was performed at the Ohio University's Edwards Accelerator Laboratory (EAL), which provides a wide range of neutron energies using \iso{Al}{27}(d,n) reactions \cite{massey1998measurement}. The 7.44~MeV deuteron beam impinges on an  $^{27}$Al target, producing neutron energy spectra at different angles relative to the beam direction. The neutron energy spectrum is well described at 120$^\circ$ relative to the beam direction and can be used as a standard spectrum to measure the efficiency of the neutron detectors. \\

   \begin{figure}[!t]
\centering
\includegraphics[width=0.47\textwidth]{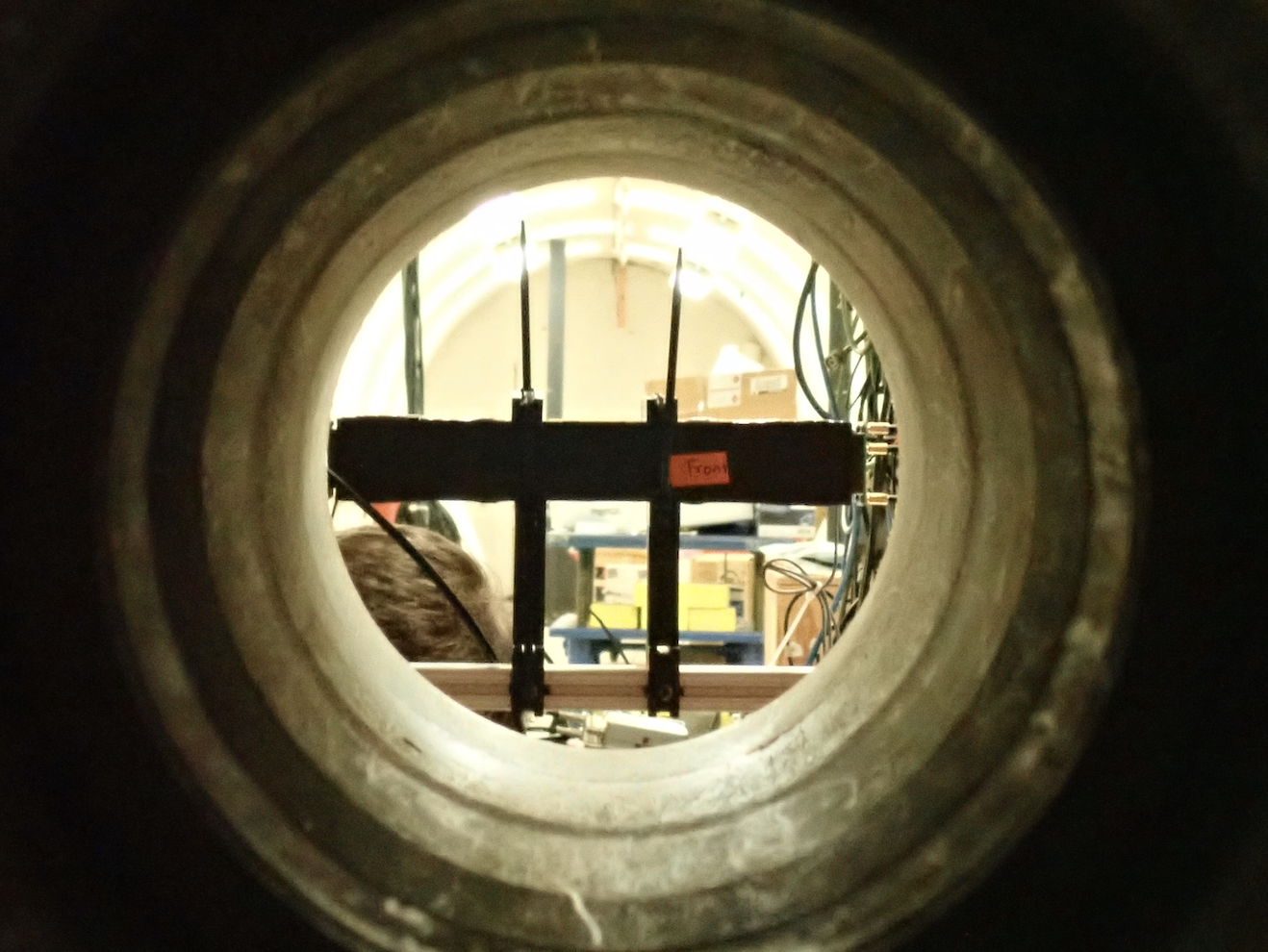}
	\caption{The experimental setup at EAL. The detector prototypes were placed at a distance of 5~m from target assembly inside the 30~m long neutron time-of-flight tunnel. The neutron beam was collimated using concentric polyethylene collimators. }
\label{fig:setup}
\end{figure}

The measurement was performed with five NEXT prototypes, four of which were constructed from EJ-276 and one from EJ-200 plastic scintillator. Four of the prototypes were 10~in long, and one of them was 5 in long. Three typical prototypes EJ276-10, EJ200-10, and EJ276-05, were considered for the analysis. 

\begin{figure}[!t]
  \centering
  \includegraphics[width=0.47\textwidth]{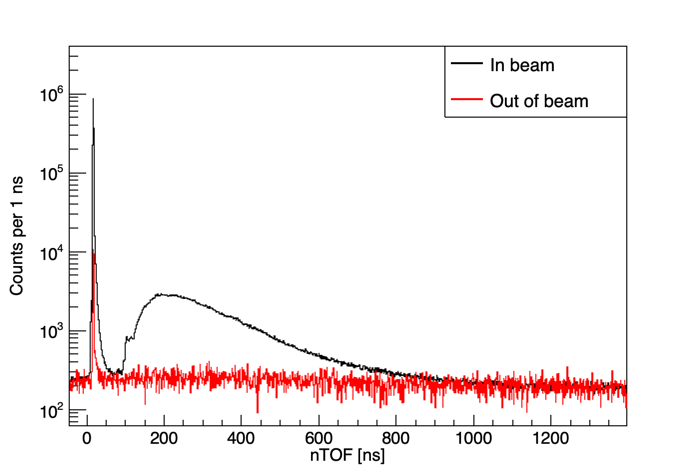}
	\caption{A typical time-of-flight spectra measured by EJ276-10 prototype with  \iso{Al}{27}(d,n) at 120$^\circ$ with prototype in beam (black) and the prototype moved out of the beam (red).}
  \label{fig:nToF}
\end{figure}

Each of the prototypes was placed at a distance of 5~m from the target assembly inside the 30~m long concrete tunnel, as shown in Figure \ref{fig:setup}. The distance was chosen such that the prototype was fully illuminated, and the neutron rate would be sufficient to make reliable measurements. A chopped pickoff signal from the beam buncher was used as a reference signal for the neutron time-of-flight measurement. Figure \ref{fig:nToF} shows a typical TOF spectrum measured by the EJ276-10 prototype in the beam, and the prototype moved out of the beam. The flat nature of the out-of-beam TOF spectrum indicates the absence of scattered neutrons inside the tunnel. The background is mainly composed of prompt and scattered $\gamma$ rays, which can be easily subtracted to obtain the actual neutron time-of-flight spectrum.

\begin{figure}[!t]
  \centering
  \includegraphics[width=0.47\textwidth]{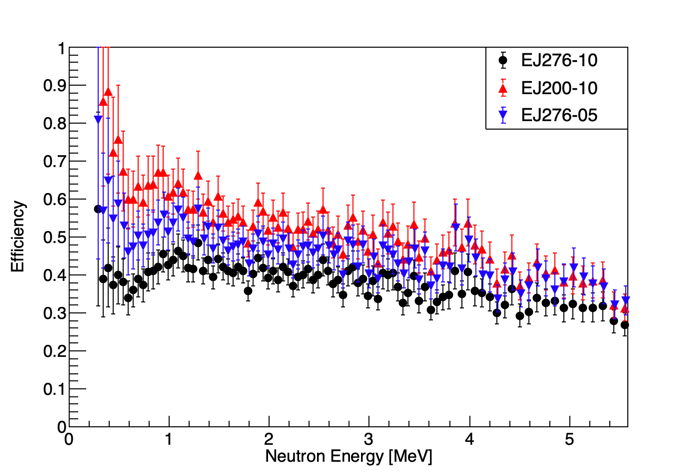}
	\caption{The efficiency for EJ276-10 (black), EJ276-05 (blue) and EJ200-10 (red) prototypes obtained from  \iso{Al}{27}(d,n) reactions at 120$^\circ$ with a 20 keVee threshold.}  \label{fig:eff_Al}
\end{figure}

For the intrinsic efficiency, the neutron flux (neutrons/sr/$\mu$C/MeV) was measured for each prototype and compared with the reference neutron spectrum provided by the facility. Figure \ref{fig:eff_Al} shows the intrinsic efficiency of three NEXT prototypes measured at EAL using the \iso{Al}{27}(d,n) reactions at 120$^\circ$. The higher light output of the EJ-200 scintillator led to the better performance of the EJ200-10 prototype for low-energy neutrons compared to the other two prototypes.

\begin{figure}[!t]
\centering
  \subfloat[]{\includegraphics[width=0.47\textwidth]{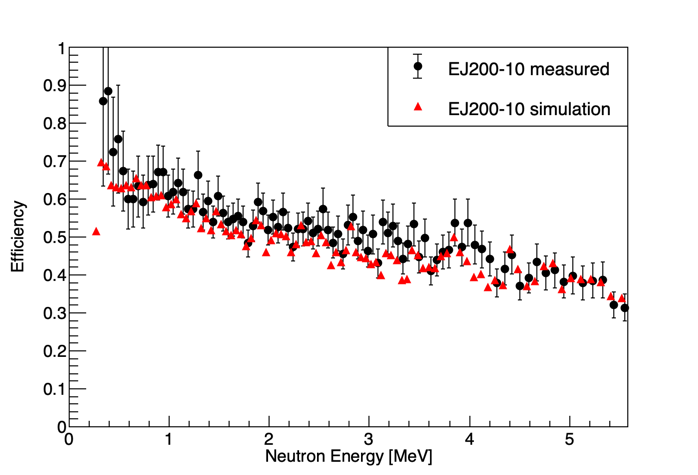} \label{fig:eff_EJ200_10in}}
  \hspace{0.3cm}
  \subfloat[]{\includegraphics[width=0.47\textwidth]{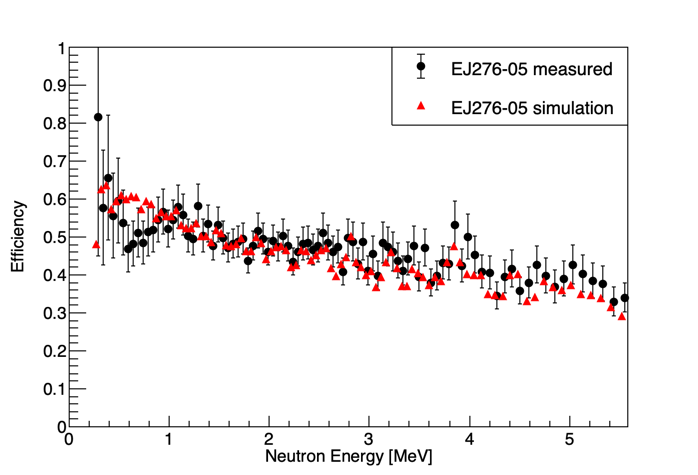} \label{fig:eff_EJ276_05in}}
    \hspace{0.3cm}
  \subfloat[]{\includegraphics[width=0.47\textwidth]{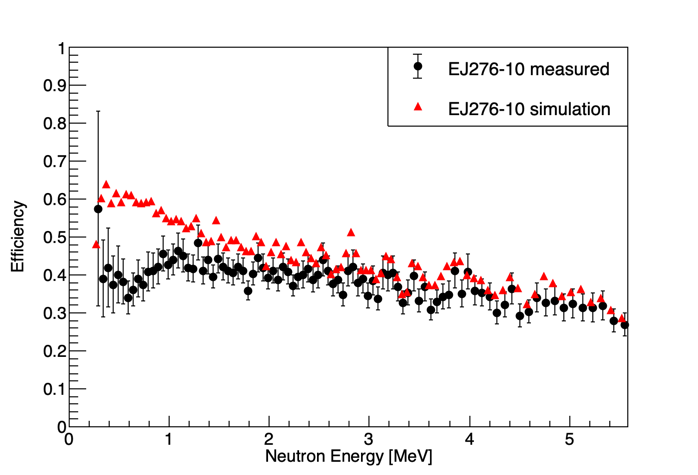} \label{fig:eff_EJ276_10in}}
   \caption{The measured intrinsic efficiency using \iso{Al}{27}(d,n) reactions at 120$^\circ$ at EAL with 20 keVee threshold (black) compared to the NEXTSim calculation (red) for EJ200-10 (a) EJ276-05 (b) and EJ276-10 (c) prototypes.}\label{fig:eff_comp}
\end{figure}

\subsection{Simulations: NEXT\emph{sim}}
NEXT\emph{sim}  \cite{heideman2019conceptual}, a GEANT4 \cite{agostinelli2003geant4, allison2016recent} based simulation framework developed to characterize the NEXT detector, was used to simulate neutron detection efficiency for the various NEXT prototypes. 

In NEXT\emph{sim}, the same experimental setup as the one used at EAL was rendered, and the same neutron flux spectrum provided by the facility was used as a source. 
Comparative studies have been performed between measured and simulated intrinsic efficiency for each prototype. 

 Figure \ref{fig:eff_comp} shows the intrinsic efficiency calculated using NEXT\emph{sim} compared to the measured efficiency for the three prototypes EJ200-10, EJ276-05, and EJ276-10. The EJ200-10 and EJ276-05 prototypes showed good agreement between measured and simulated efficiencies within the uncertainties; however, the measured intrinsic efficiency is lower than the simulated efficiency for the prototype EJ276-10 for energies below 1.5~MeV. The discrepancy is attributed to a low light yield of the EJ-276 scintillator, limitation of the Anger Logic readout, and the trigger schemes used. A new trigger configuration has been developed, leading to an improvement in detection efficiency for low-energy neutrons. The details of the readout scheme and trigger configuration are described in the next section.

\section{Anger Logic readout and trigger configurations}
 
\subsection{Readout scheme of NEXT}
 
\begin{figure*}[!t]
  \centering
  \includegraphics[width=1\textwidth]{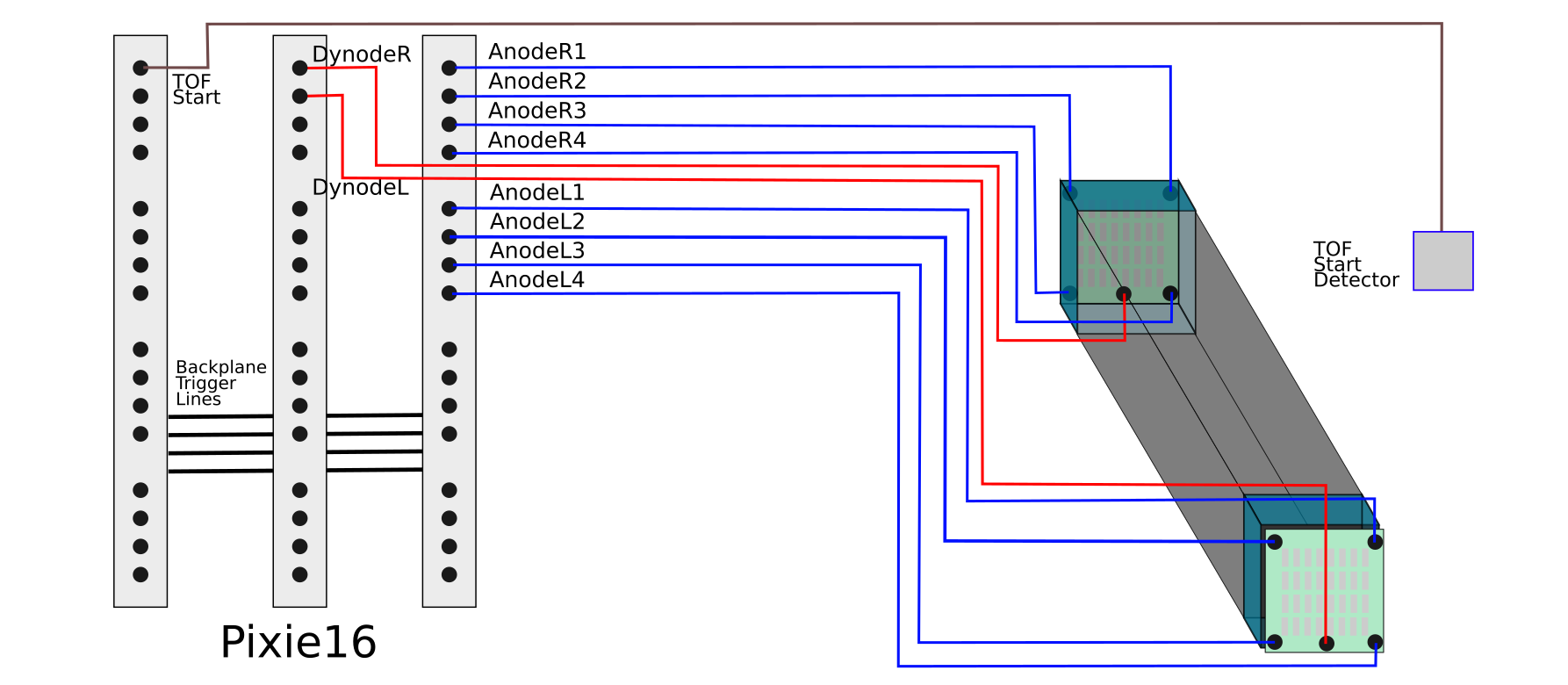}
	\caption{A schematic representation of the NEXT readout in TOF setup with start detector. The five signals from each side of the NEXT (dynode: red, anodes: blue) and a signal from the TOF start detector are digitized using Pixie16 digitizers. The TOF start detector provides a reference signal for the time-of-flight measurement. These signals communicate to each other using backplane trigger lines to generate the trigger configurations.}
  \label{fig:trigmap}
\end{figure*}

Each MAPMT used for the light readout provides one common dynode and 64 anode signals. The common dynode signal is used for timing and triggering purposes which enables an evolved implementation of the trigger scheme developed initially for VANDLE \cite{peters2016performance}, and the anode signals are used to extract the neutron interaction position in the scintillator. The dynode signal is read directly, but all 64 anodes are connected to the resistive network, and using Anger Logic, four signals are read from the corners of the resistor array, as shown schematically in Figure \ref{fig:trigmap}. Anger logic provides a cost-effective readout method of a pixelated array like NEXT because only ten electronic channels are needed to read 128 anodes and two dynodes per detector. However, there are challenges in using the resistive network readout, especially for neutrons that deposit very little energy in the scintillator. As a result of
charge division, the low energy signals can be lost when their amplitude becomes too small. This is often the case when the interaction takes place close to the
edge of the detector.

 \subsection{Trigger configurations}

\begin{figure}[!t]
  \centering
  \subfloat[]{\includegraphics[width=0.47\textwidth]{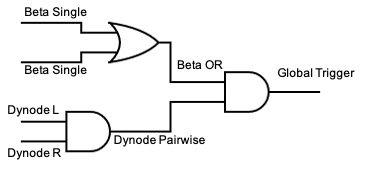}\label{fig:next_global}}
  \hspace{0.1cm}
  \subfloat[]{\includegraphics[width=0.23\textwidth]{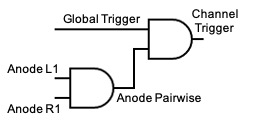}\label{fig:triples}}
  \vspace{0.05cm}
    \subfloat[]{\includegraphics[width=0.23\textwidth]{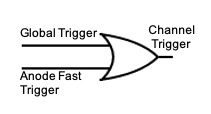}\label{fig:ftrig}}
   \caption{The schematic representation of the coincidence trigger schemes. (a) Generating the global Validation trigger using start detectors (beta singles) and NEXT dynodes. (b) A triple coincidence trigger scheme used at EAL, where anodes always required to be in coincidence with dynodes from the same detector. (c) A new trigger configuration based on unconditional readout of the anode signals. This trigger scheme allows the forced recording of a signal if global trigger is generated.}\label{fig:trigger_scheme}
\end{figure}

A coincidence triggering is required to operate the detector system with a low energy threshold without saturating the data rate. The trigger configuration used in the efficiency measurement at EAL allowed individual channel thresholds to be below noise levels. This generates two types of coincidence triggers: pairwise channel and global triggers. The pairwise triggers are generated between each channel pair (0\&1, 2\&3, ...) from the left and right sides of the detector. This approach is already advantageous in reducing detection thresholds in double-ended neutron TOF detectors. To further reject background events, a global trigger can be generated based on the TOF pairwise triggers that are coincident, within a specified time window, with channels corresponding to TOF start signals, as shown in Figure \ref{fig:next_global}.  The start signals can either be a real observation of a beta particle scattering in a plastic scintillator, a gamma from a fission source, or an RF signal tightly correlated with a beam packet arrival on a reaction target. Once a global trigger has been generated, any channel whose local fast trigger is validated alongside the global trigger records the channel event data to the FIFO.

\begin{figure}
\centering
  \subfloat[]{\includegraphics[width=0.47\textwidth]{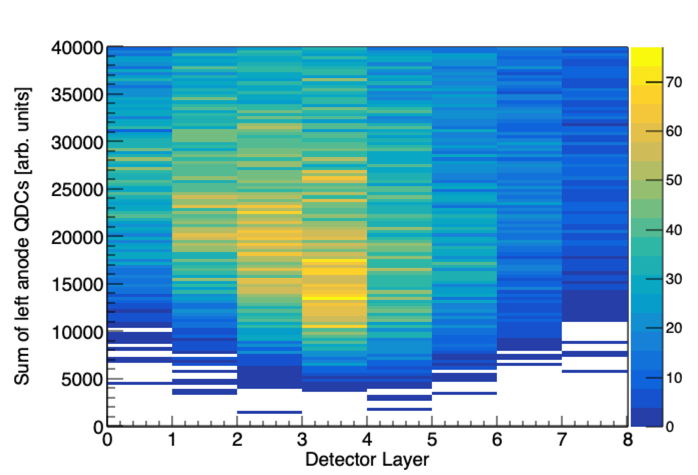}\label{fig:laqdc_triples}}\\
  \subfloat[]{\includegraphics[width=0.47\textwidth]{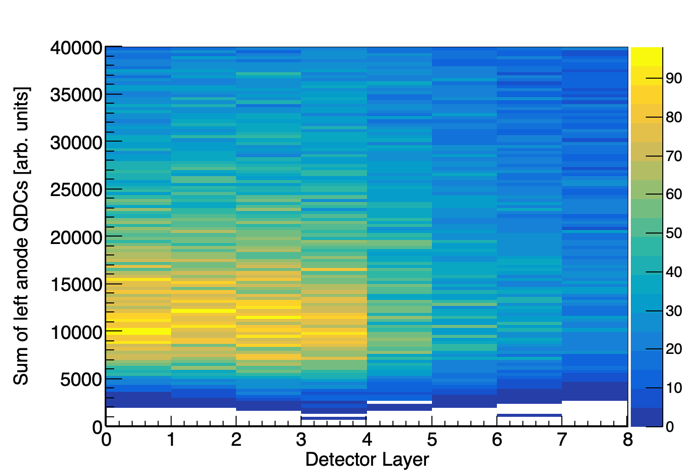}\label{fig:laqdc_ftrig}}
   \caption{Two dimensional histograms showing the sum of the left anode QDCs plotted against the corresponding detector layer for events taken with the triple coincidence trigger (a) and the new external trigger configuration (b).  The zeroth detector layer indicates the layer closest to the source. The lower occupancy in the front (towards the source) and back (away from the source) layers for low QDC events seen in the top plot does not exist in data taken with the new trigger mode. The detection threshold is consistent between the detector layers as shown in the bottom plot.}\label{fig:laqdc}
\end{figure}

Figure \ref{fig:triples} shows the schematics of the triple coincidence trigger used in the efficiency measurement at EAL.
The caveat to this trigger mode, when used with NEXT, is that the anodes, which should always be in coincidence with dynodes of the same detector, may not be recorded if the signals are too small to cross their local trigger threshold. Normally this could be remedied by simply lowering their threshold to near zero, but these anodes not only listen for global triggers but can also generate global triggers. If the anodes are set to near-zero thresholds, then too many global triggers would be generated, and the data rate would be unsustainable.
The position correction data analysis requires the four corresponding anode signals to each dynode. If the four anode signals do not exist, then the event is disregarded. For events with low energy deposition localized near the corner of the detector, the sharing of the signal through the Anger logic resistive network will
result in the opposite corner likely not triggering because its signal will be too small.
This issue resulted in a pseudo-threshold that effectively raises the detection threshold in the EAL measurement, as seen in Figure \ref{fig:laqdc_triples}.

In order to correct this, a new trigger scheme was implemented in firmware to include an unconditional readout trigger, called external fast trigger mode. This allowed the forced recording of a channel if a global trigger is generated. Anodes were set to the external fast trigger mode such that they were only recorded for globals generated between the TOF start signals and a pairwise NEXT dynode event as shown in Figure \ref{fig:ftrig}. The improvement in event construction can be seen in Figure \ref{fig:laqdc_ftrig} and will also be
demonstrated for low energy neutron detection efficiency in the next section.
\\
\\
\\

\section{\iso{Cf}{252} measurement with new trigger scheme } 
A measurement of the \iso{Cf}{252} fission neutron spectrum was performed to show the improvement in the low-energy neutron detection using a new trigger scheme compared to the trigger scheme used during efficiency measurement at EAL.

The measurement was performed using both the old triple coincidence trigger and the new external fast trigger configurations in the same experimental setup. The same prototype (EJ276-10) from the EAL measurement was used and placed at a distance of 94.4 cm from the source. The start signal for the TOF measurement was provided by detecting prompt $\gamma-$rays from fission in a 25.4$\times$25.4$\times$12.7~mm\textsuperscript{3} plastic scintillator (EJ-200) placed close to the source and the NEXT prototype provided the stop signal. A shadow bar measurement was also performed in a similar setup to see the background neutrons scattered off the floor, walls, and the surrounding materials. The blocks of polyethylene cubes inserted between the source and the NEXT prototype blocked direct neutrons from the source. After subtracting the background, a clean time-of-flight spectrum was obtained, and corresponding neutron energies were extracted.

\begin{figure}[!t]
\centering
  \subfloat[]{\includegraphics[width=0.47\textwidth]{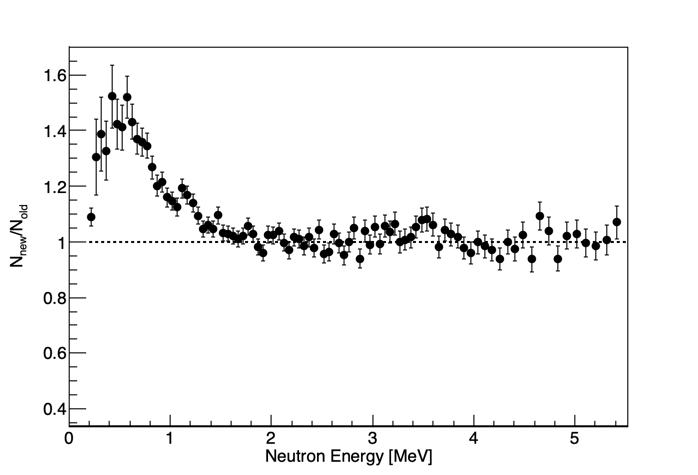}\label{fig:ratio}}
  \hspace{0.3cm}
  \subfloat[]{\includegraphics[width=0.47\textwidth]{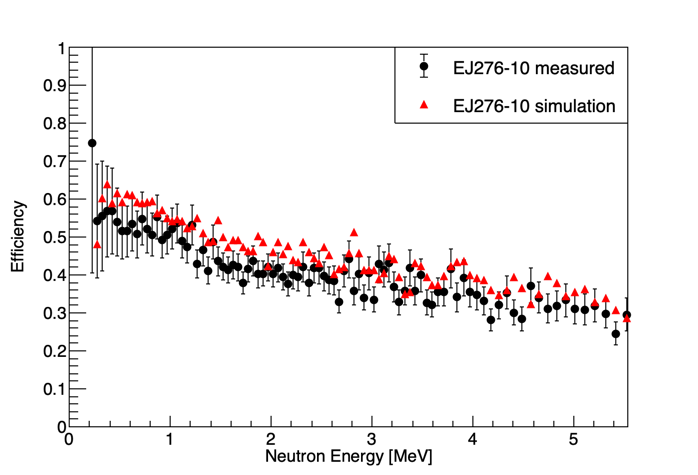}\label{fig:corrected_eff}}
   \caption{ (a) Ratio of number of neutrons measured per energy bin in new and old trigger schemes with the EJ276-10 prototype using \iso{Cf}{252} source. (b) The efficiency for EJ276-10 prototype measured at EAL was corrected using the ratio of neutrons detected in new and old trigger scheme shown in Figure \ref{fig:ratio}.}\label{fig:eff_corr}
\end{figure}

Figure \ref{fig:ratio} shows the ratio of the number of neutrons detected per energy bin in the new and old trigger configurations. With the new trigger configuration, the detection efficiency for the low-energy neutrons (below 1.5 MeV) is higher than the old one. This removed the discrepancy seen in Figure \ref{fig:eff_EJ276_10in} between the measured and simulated efficiency for the EJ276-10 prototype, as shown in Figure \ref{fig:corrected_eff}. The efficiency measured at EAL was multiplied by the ratio of the neutron detected in the new and old trigger scheme to get the corrected efficiency.

\section{Updates in the detector development}

Updates have been made in the detector development since the EAL efficiency measurement. The Anger Logic readout boards now have onboard pre-amplifiers, which will be used to amplify both dynode and anode signals. The scintillator plastic to be used will also be changed to EJ-299-33M as recommended
by the supplier (Eljen Technologies). This plastic has a higher light output and similar
n-$\gamma$ discrimination compared to the EJ-276 used in measurements shown in this
work. These updates in the detector development will improve the neutron detection capabilities. Future publications will include the results from the characterization of the new detector units.

\section{Conclusion}
We have demonstrated a high neutron detection efficiency of NEXT prototypes using neutrons produced from the \iso{Al}{27}(d,n) reactions. Efficiency measurements of three typical NEXT prototypes, which differ in length and the scintillator material, were performed and analyzed. The performance of each of the prototypes is compatible with the prediction from GEANT4 simulations. A new trigger configuration has been developed, which leads to better measurements of low-energy neutrons.   New developments have been made in the readout scheme, scintillator material, and the detector manufacturing procedure. These updates will improve the neutron detection capabilities and the neutron detection efficiency of the NEXT detector. 


\section*{Acknowledgments}
This work was supported by the U.S. Department of Energy, National Nuclear Security Administration under the Stewardship Science Academic Alliances program through DOE Award No. DE-NA0002934 and DE-NA0003899, NSF Major Research Instrumentation Program Award Number 1919735, and DOE Award No. DE-FG02-96ER4093. The Tennessee Technological University team was partially supported through a grant.
No. DE-SC0016988.

\label{}

 \bibliography{next_efficiency}

\begin{thebibliography}{10}
\expandafter\ifx\csname url\endcsname\relax
  \def\url#1{\texttt{#1}}\fi
\expandafter\ifx\csname urlprefix\endcsname\relax\def\urlprefix{URL }\fi
\expandafter\ifx\csname href\endcsname\relax
  \def\href#1#2{#2} \def\path#1{#1}\fi

\bibitem{dimitriou2021development}
P.~Dimitriou, I.~Dillmann, B.~Singh, V.~Piksaikin, K.~Rykaczewski, J.~Tain,
  A.~Algora, K.~Banerjee, I.~Borzov, D.~Cano-Ott, et~al., Development of a
  reference database for beta-delayed neutron emission, Nuclear Data Sheets 173
  (2021) 144--238.

\bibitem{nakamura2017exotic}
T.~Nakamura, H.~Sakurai, H.~Watanabe, Exotic nuclei explored at in-flight
  separators, Progress in Particle and Nuclear Physics 97 (2017) 53--122.

\bibitem{pfutzner2012radioactive}
M.~Pf{\"u}tzner, M.~Karny, L.~Grigorenko, K.~Riisager, Radioactive decays at
  limits of nuclear stability, Reviews of modern physics 84~(2) (2012) 567.

\bibitem{roberts1939further}
R.~Roberts, R.~Meyer, P.~Wang, Further observations on the splitting of uranium
  and thorium, Physical Review 55~(5) (1939) 510.

\bibitem{doi:10.1080/10619127.2013.855558}
A.~Gade, C.~K. Gelbke, T.~Glasmacher,
  \href{https://doi.org/10.1080/10619127.2013.855558}{Nscl and the facility for
  rare isotope beams (frib) project}, Nuclear Physics News 24~(1) (2014)
  28--30.
\newblock \href
  {http://arxiv.org/abs/https://doi.org/10.1080/10619127.2013.855558}
  {\path{arXiv:https://doi.org/10.1080/10619127.2013.855558}}, \href
  {https://doi.org/10.1080/10619127.2013.855558}
  {\path{doi:10.1080/10619127.2013.855558}}.
\newline\urlprefix\url{https://doi.org/10.1080/10619127.2013.855558}

\bibitem{heideman2019conceptual}
J.~Heideman, D.~P{\'e}rez-Loureiro, R.~Grzywacz, C.~Thornsberry, J.~Chan,
  L.~Heilbronn, S.~Neupane, K.~Schmitt, M.~Rajabali, A.~Engelhardt, et~al.,
  Conceptual design and first results for a neutron detector with interaction
  localization capabilities, Nuclear Instruments and Methods in Physics
  Research Section A: Accelerators, Spectrometers, Detectors and Associated
  Equipment 946 (2019) 162528.

\bibitem{eljen}
Eljen technologies, \url{https://www.eljentechnology.com}.

\bibitem{3MESR}
{3M\textsuperscript{TM}}, \url{https://www.3m.com/}.

\bibitem{HMPMT}
{Hamamatsu Photonics K. K.}, \url{https://www.hamamatsu.com/jp/en/index.html}.

\bibitem{VERTILON}
{Vertilon Corporation, Westford, MA}, \url{https://www.vertilon.com/}.

\bibitem{anger1964scintillation}
H.~O. Anger, Scintillation camera with multichannel collimators (1964).

\bibitem{pixie16}
{XIA LLC, 31057 Genstard Rd., Hayward, CA}, \url{https://www.xia.com/}.

\bibitem{massey1998measurement}
T.~Massey, S.~Al-Quraishi, C.~Brient, J.~Guillemette, S.~Grimes, D.~Jacobs,
  J.~O'Donnell, J.~Oldendick, R.~Wheeler, A measurement of the 27al (d, n)
  spectrum for use in neutron detector calibration, Nuclear science and
  engineering 129~(2) (1998) 175--179.

\bibitem{agostinelli2003geant4}
S.~Agostinelli, J.~Allison, K.~a. Amako, J.~Apostolakis, H.~Araujo, P.~Arce,
  M.~Asai, D.~Axen, S.~Banerjee, G.~. Barrand, et~al., Geant4---a simulation
  toolkit, Nuclear instruments and methods in physics research section A:
  Accelerators, Spectrometers, Detectors and Associated Equipment 506~(3)
  (2003) 250--303.

\bibitem{allison2016recent}
J.~Allison, K.~Amako, J.~Apostolakis, P.~Arce, M.~Asai, T.~Aso, E.~Bagli,
  A.~Bagulya, S.~Banerjee, G.~Barrand, et~al., Recent developments in geant4,
  Nuclear Instruments and Methods in Physics Research Section A: Accelerators,
  Spectrometers, Detectors and Associated Equipment 835 (2016) 186--225.

\bibitem{peters2016performance}
W.~Peters, S.~Ilyushkin, M.~Madurga, C.~Matei, S.~Paulauskas, R.~Grzywacz,
  D.~Bardayan, C.~Brune, J.~Allen, J.~Allen, et~al., Performance of the
  versatile array of neutron detectors at low energy (vandle), Nuclear
  Instruments and Methods in Physics Research Section A: Accelerators,
  Spectrometers, Detectors and Associated Equipment 836 (2016) 122--133.

\end{thebibliography}

\end{document}